\documentclass[
reprint,           
superscriptaddress,
amsmath,           
amssymb,           
aps,               
prd,               
notitlepage,       
floatfix,          
nofootinbib 
]{revtex4-1}

\usepackage{tensor}     
\usepackage{graphicx}   
\usepackage[
colorlinks=true,        
citecolor=blue,         
linkcolor=blue,         
urlcolor=blue           
]{hyperref}             
\usepackage{bm}         
\usepackage{xcolor}     
\usepackage{tabulary}
\usepackage{color}      
\usepackage[utf8]{inputenc} 
\usepackage[section]{placeins} 
\usepackage{ulem}
\usepackage{makecell}
\usepackage{multirow}
\newcommand{\nc}{\newcommand*} 

\nc{\al}{\alpha}
\nc{\s}{\sigma}
\nc{\dt}{\delta}
\nc{\Dt}{\Delta}
\nc{\Ld}{\Lambda}
\nc{\p}{\partial}
\nc{\om}{\omega}
\nc{\Om}{\Omega}
\nc{\rd}{\mathrm{d}}
\nc{\Od}[1]{\mathcal{O}(#1)} 
\nc{\kp}{\kappa}
\nc{\one}{\uppercase\expandafter{\romannumeral1}}
\nc{\two}{\uppercase\expandafter{\romannumeral2}}
\nc{\three}{\uppercase\expandafter{\romannumeral3}}
\def\({\left(}
\def\){\right)}
\def\[{\left[}
\def\]{\right]}
\def\e{\begin{equation}}
\def\q{\end{equation}}
\def\m{\begin{eqnarray}}
\def\n{\end{eqnarray}}
\nc{\Eq}[1]{Eq.~\eqref{#1}}     
\nc{\Fig}[1]{Fig.~\ref{#1}}     
\nc{\Table}[1]{Table~\ref{#1}}  
\nc{\Sec}[1]{Sec.~\ref{#1}}     
\nc{\Msun}{M_\odot}             
\nc{\fpbh}{f_{\mathrm{pbh}}}    
\nc{\fpbhn}{f_{\mathrm{pbh0}}}    
\nc{\mR}{\mathcal{R}} 
\nc{\seq}{\sigma_{\mathrm{eq}}}
\nc{\ogw}{\Omega_{\mathrm{GW}}}
\nc{\gpcyr}{\mathrm{Gpc}^{-3}\,\mathrm{yr}^{-1}}
\nc{\lvc}{LIGO/Virgo} 
\nc{\SNR}{\mathrm{SNR}} 
\nc{\mmin}{{m_{\mathrm{min}}}}
\nc{\mmax}{{m_{\mathrm{max}}}}
\nc{\Mmin}{{M_{\mathrm{min}}}}
\nc{\fmin}{{f_{\mathrm{min}}}}
\nc{\VT}{\mathrm{VT}}
\nc{\rhoGW}{\rho_{\mathrm{GW}}}
\nc{\vth}{\vec{\theta}}
\nc{\vd}{\vec{d}}
\nc{\vla}{\vec{\lambda}}
\nc{\Nobs}{N_{\mathrm{obs}}}
\nc{\av}[1]{\langle #1 \rangle} 
\nc{\km}{\mathrm{km}}
\nc{\Mpc}{\mathrm{Mpc}}
\nc{\Tobs}{T_{\mathrm{obs}}}
\nc{\Ntemp}{N_{\mathrm{temp}}}
\nc{\addref}{[\textcolor{red}{add ref}] } 
\nc{\eg}{\textit{e.g.~}}
\nc{\app}{\approx}
\nc{\hf}{\frac{1}{2}}
\nc{\discuss}{\textcolor{red}{Add discussion here!}}
\nc{\red}[1]{\textcolor{red}{#1}}
\nc{\mH}{\mathcal{H}}
\nc{\cs}{c_s^2}
\nc{\Sij}[1]{S_{ij}^{(#1)}}
\nc{\vi}[1]{v_i^{(#1)}}
\nc{\no}{\nonumber}
\def\<{\left\langle}
\def\>{\right\rangle}

\def\half{{1\over 2}}
\nc{\bk}{\bm{k}}
\nc{\bq}{\bm{q}}
\nc{\bp}{\bm{p}}
\nc{\bl}{\bm{l}}
\nc{\bx}{\bm{x}}
\nc{\be}{\mathbf{e}}
\nc{\mS}{\mathcal{S}}
\nc{\te}{\tilde{\eta}}
\nc{\tp}{\tilde{p}}
\nc{\tk}{\tilde{k}}
\nc{\tx}{\tilde{x}}
\nc{\tF}{\tilde{F}}
\nc{\tA}{\tilde{A}}
\nc{\mkpq}{|\bk-\bp-\bq|}
\nc{\mpq}{|\bp-\bq|}
\nc{\mkp}{|\bk-\bp|}
\nc{\mSi}[1]{\mS^{(#1)}({\bk, \eta})}
\nc{\vk}{\vec{k}}
\nc{\kstar}{k_*}
\nc{\fstar}{f_*}
\nc{\xstar}{x_*}
\nc{\mpbh}{m_{\rm{pbh}}}
\nc{\bn}[1]{\bm{n}_{\text{#1}}}
\nc{\bC}[1]{\bm{C}_{\text{#1}}}
\nc{\NTOA}{N_{\text{TOA}}}
\nc{\Nmode}{{N_{\text{mode}}}}
\nc{\ARN}{A_{\rm{RN}}}
\nc{\gRN}{\gamma_{\rm{RN}}}
\nc{\bS}{\mathbf{\Sigma}}
\nc{\br}{\mathbf{r}}
\nc{\bN}{\mathbf{N}}
\nc{\fnl}{F_{\mathrm{NL}}}
\nc{\gnl}{G_{\mathrm{NL}}}

\renewcommand{\vec}[1]{\boldsymbol{#1}} 

\nc{\arXiv}[2]{\href{http://arxiv.org/pdf/#1}{{\tt [#2/#1]}}}
\nc{\arXivold}[1]{\href{http://arxiv.org/pdf/#1}{{\tt [#1]}}}

\renewcommand{\vec}[1]{\boldsymbol{#1}} 

\begin{document}
	


\title{One-loop correction to the  enhanced curvature perturbation with local-type non-Gaussianity for the formation of primordial black holes}
	
\author{De-Shuang Meng}
\email{mengdeshuang@itp.ac.cn}
\affiliation{CAS Key Laboratory of Theoretical Physics,
	Institute of Theoretical Physics, Chinese Academy of Sciences,
	Beijing 100190, China}
\affiliation{School of Physical Sciences,
	University of Chinese Academy of Sciences,
	No. 19A Yuquan Road, Beijing 100049, China}
\author{Chen Yuan}
\email{Corresponding author: yuanchen@itp.ac.cn}
\affiliation{CAS Key Laboratory of Theoretical Physics,
Institute of Theoretical Physics, Chinese Academy of Sciences,
Beijing 100190, China}
\affiliation{School of Physical Sciences,
University of Chinese Academy of Sciences,
No. 19A Yuquan Road, Beijing 100049, China}
\author{Qing-Guo Huang}
\email{Corresponding author: huangqg@itp.ac.cn}
\affiliation{CAS Key Laboratory of Theoretical Physics, 
		Institute of Theoretical Physics, Chinese Academy of Sciences,
		Beijing 100190, China}
\affiliation{School of Physical Sciences, 
		University of Chinese Academy of Sciences, 
		No. 19A Yuquan Road, Beijing 100049, China}
\affiliation{School of Fundamental Physics and Mathematical Sciences
		Hangzhou Institute for Advanced Study, UCAS, Hangzhou 310024, China}

\date{\today}

\begin{abstract}
As one of the promising candidates of cold dark matter (DM), primordial black holes (PBHs) were formed due to the collapse of over-densed regions generated by the enhanced curvature perturbations during the radiation-dominated era. The enhanced curvature perturbations are expected to be non-Gaussian in some relevant inflation models and hence the  higher-order loop corrections to the curvature power spectrum might be non-negligible as well as altering the abundance of PBHs. In this paper, we calculate the one-loop correction to the curvature power spectrum with  local-type non-Gaussianities characterizing by $F_{\mathrm{NL}}$ and $G_{\mathrm{NL}}$ standing for the quadratic and cubic non-Gaussian parameters, respectively. Requiring that the one-loop correction be subdominant, we find a perturbativity condition, namely $|2cAF_{\mathrm{NL}}^2+6AG_{\mathrm{NL}}|\ll 1$, where $c$ is a constant coefficient which can be explicitly calculated in the given model and $A$ denotes the variance of  Gaussian part of enhanced curvature perturbation, and such a perturbativity  condition can provide a stringent constraint on the relevant inflation models for the formation of PBHs. 

\end{abstract}
	
\pacs{???}
	
\maketitle

\section{Introduction}

Primordial black holes (PBHs) can form from the collapse of over-densed regions when large curvature perturbations re-enter the horizon during the radiation dominated (RD) era \cite{Zeldovich:1967lct,Hawking:1971ei,Carr:1974nx,Carr:1975qj}. PBHs can not only represent cold dark matter (DM) but also explain the merger events detected by LIGO-Virgo Collaboration \cite{Sasaki:2016jop,Chen:2018czv,Raidal:2018bbj,DeLuca:2020qqa,Hall:2020daa,Bhagwat:2020bzh,Hutsi:2020sol,Wong:2020yig,DeLuca:2021wjr,Franciolini:2021tla,Franciolini:2021tla,Chen:2021nxo}.
There are various investigations \cite{Carr_2010,Chen_2016,Ali_Ha_moud_2017,Aloni_2017,https://doi.org/10.48550/arxiv.1612.07264,Bird:2016dcv,Garcia-Bellido:2017fdg,Sasaki:2018dmp,Barack:2018yly,Chen:2018czv,Chen:2018rzo,Chen:2019irf,Barnacka_2012,Graham:2015apa,Niikura:2017zjd,PhysRevLett.111.181302,EROS-2:2006ryy,Brandt:2016aco,Gaggero:2016dpq,Niikura:2019kqi,Wang:2016ana,LIGOScientific:2018glc,Magee:2018opb,Montero-Camacho:2019jte,Laha:2019ssq,Chen:2019xse,Chen:2021nxo} putting constraints on the fraction of PBHs in DM to no more than a few percent except two mass windows $\[10^{-16},10^{-14}\]\Msun \cup \[10^{-13},10^{-12}\] \Msun$ \cite{Defillon_2014,Katz_2018}. Review of constraints on PBHs can be found in \cite{Carr:2020gox,Carr:2020xqk}.

It is estimated that the curvature power spectrum needs to be enhanced to about $10^{-2}$ in order to form sufficient PBHs on certain small scales, compared to those on the cosmic microwave background (CMB) scales, which is of order $10^{-9}$ \cite{Planck:2018jri}.
PBHs are formed at the tail of the probability density function (PDF) of the curvature perturbations, hence the formation of PBHs are extremely sensitive to non-Gaussianities.
On the other hand, in the squeezed limit of the bispectrum for single field inflation models, the Maldacena consistency condition \cite{Maldacena:2002vr} for the non-Gaussian parameter $f_{\text{NL}}^{\text{sq}}$ and the spectral index $n_s$, namely $f_{\text{NL}}^{\text{sq}} = -5(n_s-1)/12$, is expected to hold. Therefore non-negligible non-Gaussianities are usually accompanied by the enhancement of power spectrum where the spectral index would be much larger, rendering non-Gaussianities might play an significant role in those inflation models that predict large numbers of PBHs.

From the viewpoint of quantum field theory (QFT), if one takes the interaction picture, the power spectrum of curvature perturbations is equivalent to calculate the vacuum expectation value of two-point correlation function (2PCF) and the  non-Gaussianities correspond to N-point correlation function (NPCF) with $N>2$ by in-in formalism. On the other hand, the NPCF can make contribution to the 2PCF through loop corrections. Based on the fact that loop corrections of 2PCF need to be smaller than the tree level in order to maintain the significance of perturbation theory, the authors in  \cite{Kristiano:2021urj} found a perturbativity condition such that $c_s^{-4} \mathcal{P}_{\zeta} \ll 1-n_s$ for a single-field inflation model, where $c_s$ represents the sound speed, $\mathcal{P}_{\zeta}$ denotes the amplitude of the tree level power spectrum and $n_s<1$ is the spectral index.

In this paper, we will  calculate the one-loop correction to the curvature power spectrum with local-type non-Gaussianities and work out the perturbativity condition for the enhanced curvature perturbation for the formation of PBHs. 
Besides the constraints from loop corrections, the abundance of PBHs would naturally select the non-Gaussian parameters (see e.g., \cite{Byrnes:2012yx}). And then we investigate the constraints on non-Gaussian parameters by taking into account both the perturbativity condition and PBHs abundance. The paper will be organized as follows. In Sec.~{\ref{oneloop}}, we calculate the one-loop correction for enhanced power spectra of curvature perturbations from local-type non-Gaussianities, and work out the perturbativity condition. In Sec.~{\ref{abundance}}, we review the calculation of PBH abundance and obtain the constraints on the non-Gaussian parameters. Finally, we give a brief conclusion and discussion in Sec.~\ref{cd}.

 
\section{Constraints on the local-type non-Gaussian parameters from one-loop corrections}
\label{oneloop}

For the local-type non-Gaussianities, the curvature perturbation is expanded in terms of the Gaussian part in real space. Up to cubic order, it is  given by
\m\label{local}
\zeta(\vec{x})=\zeta_\text{g}(\vec{x}) + \fnl \zeta_\text{g}^2(\vec{x}) + \gnl \zeta_\text{g}^3(\vec{x}),
\n
where $\zeta_\text{g}(\vec{x})$ follows Gaussian statistics and $\fnl$ and $\gnl$ are the dimensionless non-Gaussian parameters, related to the commonly used notations $f_{\mathrm{NL}}$ and $g_{\mathrm{NL}}$ by $\fnl \equiv 3/5 f_{\mathrm{NL}}$ and $\gnl \equiv 9/25 g_{\mathrm{NL}}$ respectively. In momentum space, the curvature perturbation is expanded by convolution of the Gaussian part
\m\label{loc}
\zeta(\vk) =&& \zeta_{\text{g}}(\vk) + F_{\mathrm{NL}} \int \frac{\mathrm{d}^{3} p}{(2 \pi)^{3}} \zeta_{\text{g}}(\vec{p}) \zeta_{\text{g}}(\vec{k}-\vec{p}) \no\\
&& + G_{\mathrm{NL}} \int \frac{\mathrm{d}^{3} p}{(2 \pi)^{3}} \int \frac{\mathrm{d}^{3} q}{(2 \pi)^{3}} \zeta_{\text{g}}(\vec{p}) \zeta_{\text{g}}(\vec{q}) \zeta_{\text{g}}(\vec{k}-\vec{p}-\vec{q}). \no\\
\n
The dimensionless power spectrum of curvature perturbation,  $\mathcal{P}_\zeta(k)$, is defined as
\m\label{ps}
\< \zeta(\vk) \zeta(\vk^{\prime}) \> = \frac{2 \pi^{2}}{k^{3}} \mathcal{P}_\zeta(k) (2 \pi)^{3} \delta^{(3)} (\vk + \vk^{\prime}).
\n
The one-loop correction from the local-type non-Gaussianities can be derived by inserting Eq.~(\ref{loc}) into \Eq{ps}. According to the property of a Gaussian variable, the odd $n$-point functions vanish and the even $n$-point functions can be expanded by all possible contractions of the 2PCFs, $\< \zeta_{\text{g}}(\vk) \zeta_{\text{g}}(\vk^{\prime}) \>$, and the final result can be expressed as
\m
\< \zeta(\vk) \zeta(\vk^{\prime}) \> = && \< \zeta_{\text{g}}(\vk) \zeta_{\text{g}}(\vk^{\prime}) \> + \int \frac{\rd^{3} p}{(2 \pi)^{3}} \int \frac{\rd^3 q}{(2 \pi)^{3}} \no\\ 
&& \times \big[ F_{\mathrm{NL}}^2 \< \zeta_{\text{g}}(\vec{p}) \zeta_{\text{g}}(\vk-\vec{p}) \zeta_{\text{g}}(\vec{q}) \zeta_{\text{g}}(\vk^{\prime}-\vec{q}) \> \no\\
&& + 2 G_{\mathrm{NL}} \< \zeta_{\text{g}}(\vk) \zeta_{\text{g}}(\vec{p}) \zeta_{\text{g}}(\vec{q}) \zeta_{\text{g}}(\vec{k}^{\prime}-\vec{p}-\vec{q}) \> \big] \no\\
= && \< \zeta_{\text{g}}(\vk) \zeta_{\text{g}}(\vk^{\prime}) \> + \int \frac{\rd^3 p}{(2 \pi)^{3}} \int \frac{\rd^{3} q}{(2 \pi)^{3}} \no\\
&& \times \big[ 2 F_{\text{NL}}^2 \< \zeta_{\text{g}}(\vec{p}) \zeta_{\text{g}}(\vec{q}) \> \< \zeta_{\text{g}}(\vk-\vec{p}) \zeta_{\text{g}}(\vk^{\prime}-\vec{q}) \> \no\\
&& + 6 G_{\mathrm{NL}} \< \zeta_{\text{g}}(\vk) \zeta_{\text{g}}(\vec{p}) \> \< \zeta_{\text{g}}(\vec{q}) \zeta_{\text{g}}(\vec{k}^{\prime}-\vec{p}-\vec{q}) \> \big], \no\\
\n
where the disconnected diagram vanishes. Therefore, the dimensionless power spectrum can be written as
\e
\mathcal{P}_\zeta(k) = \mathcal{P}^{(0)}_\zeta(k) + \mathcal{P}^{(1)}_\zeta(k),
\q
where $\mathcal{P}^{(0)}_\zeta(k)$ is the Gaussian part spectrum and
\m\label{1ls}
\mathcal{P}^{(1)}_\zeta(k) = && \frac{k^3 F_{\text{NL}}^2}{2 \pi} \int \rd^{3} p \frac{\mathcal{P}^{(0)}_\zeta(p) \mathcal{P}^{(0)}_\zeta( \left| \vk - \vec{p} \right|)}{p^{3} \left|\vk - \vec{p} \right|^3} \no\\
&& + \frac{3 G_{\mathrm{NL}}}{2 \pi} \mathcal{P}^{(0)}_\zeta(k) \int \mathrm{d}^{3} p \frac{\mathcal{P}^{(0)}_\zeta(p)}{p^{3}}
\n
is the one-loop correction. By introducing two variables $u = p/k $ and $v = |\vec{k}-\vec{p}|/k $, the one-loop correction can be rewritten as
\m\label{1loop}
\mathcal{P}_\zeta ^{(1)}(k) = && F_{\text{NL}}^2 \int_0^{\infty} \mathrm{d} u \int_{|1-u|}^{1+u} \mathrm{d} v \frac{\mathcal{P}^{(0)}_\zeta(uk) \mathcal{P}^{(0)}_\zeta(vk)}{u^2v^2} \no\\
&& + 6G_{\mathrm{NL}} \mathcal{P}^{(0)}_\zeta(k) \int \mathrm{d} p \frac{ \mathcal{P}^{(0)}_\zeta(p)}{p}\no\\
= && F_{\text{NL}}^2 \int_0^{\infty} \mathrm{d} u \int_{|1-u|}^{1+u} \mathrm{d} v \frac{\mathcal{P}^{(0)}_\zeta(uk) \mathcal{P}^{(0)}_\zeta(vk)}{u^2v^2} \no\\
&& + 6 A G_{\mathrm{NL}} \mathcal{P}^{(0)}_\zeta(k).
\n
where $A$ stands for the variance of the Gaussian part of curvature perturbation  spectrum,  $\mathcal{P}^{(0)}_\zeta(k)$, namely 
\e
A = \int \mathcal{P}^{(0)}_\zeta(k) \mathrm{d} \ln k.
\q
And the variance of the one-loop correction $\mathcal{P}_\zeta ^{(1)}(k)$ reads
\m\label{sigma1}
\sigma^{(1)} = && \int \mathcal{P} ^{(1)}_{\zeta}(k) \mathrm{d} \ln k \no \\
= && F_{\text{NL}}^2 \int \mathrm{d} \ln k \int_{0}^{\infty} \mathrm{d} u \int_{|1-u|}^{1+u} \mathrm{d} v \frac{\mathcal{P}^{(0)}_\zeta(uk) \mathcal{P}^{(0)}_\zeta(vk)}{u^2v^2} \no\\
&& + 6 A^2 \gnl \no\\
= && 2cA^2 F_{\text{NL}}^2+6 A^2 \gnl,
\label{sigma1}
\n
where $c$ is a constant coefficient which can be explicitly calculated for the given enhanced curvature perturbation. Usually $c$ is expected to be order of ${\cal O}(1)$ for some typical models.
From the viewpoint of quantum field theory (QFT), $\fnl$ and $\gnl$ should be regarded as the coupling constants. In general, the shape of the power spectrum from one-loop correction should be different from that in tree-level order.  Quantitatively, the variance of the one-loop correction, $\sigma^{(1)}$, is supposed to be much smaller than that of the tree-level order, $\sigma^{(0)} = A$, in order to ensure the expansion converges. Therefore the perturbativity condition for the enhanced curvature perturbation with local-type non-Gaussianities reads 
\m
|2cAF_{\mathrm{NL}}^2+6AG_{\mathrm{NL}}|\ll 1. 
\n
Note that $\gnl$ can be positive or negative. 


In the following part of this section, we will consider two typical models that are enhanced at a certain scale over the CMB scale, namely an infinite narrow spectrum and a log-normal shape spectrum. These two models are commonly used in studying the formation of PBHs (see e.g., \cite{Kohri:2018awv,Bartolo:2018rku,Bartolo:2018evs,Yuan:2019udt,Chen:2019xse}). Suppose that an enhanced power spectrum has a cut-off at $k_{\max}$ (or decrease dramatically if $k>k_{\max}$), then according to Eq.~(\ref{1ls}), the one-loop power spectrum will have a cut-off at $2k_{\max}$, due to the conservation of momentum. This indicates that, for an enhanced power spectrum which has a cut-off wavelength, the one-loop power spectrum is integrable and one does not need to perform regularization and renormalization which is different to the case discussed in \cite{Kristiano:2021urj} where the authors consider a scale-invariant power spectrum.

The infinite narrow spectrum peaked at $k_*$ at the tree level, namely the $\delta$ spectrum, is parameterized as
\e\label{delta}
\mathcal{P}_\zeta^{(0)}(k) = A k_* \delta^{(3)}\(k-k_*\).
\q
Then the one-loop correction to the $\delta$ power spectrum for local non-Gaussian expansion can be analytically expressed by 
\m
\mathcal{P}_\zeta ^{(1)}(k) =&& A^2 F_{\text{NL}}^2 \(\frac{k}{k_*}\)^2 \Theta\(2-\frac{k}{k_*}\)\no\\
&& + 6 A^2 G_{\text{NL}} k_*\delta\(k-k_*\),
\n
and the variance of the one-loop correction is $\sigma^{(1)}=2A^2\fnl^2+6A^2\gnl$, corresponding to $c=1$ in Eq.~(\ref{sigma1}). 

\begin{figure}
	\centering
	\includegraphics[width=0.9\columnwidth]{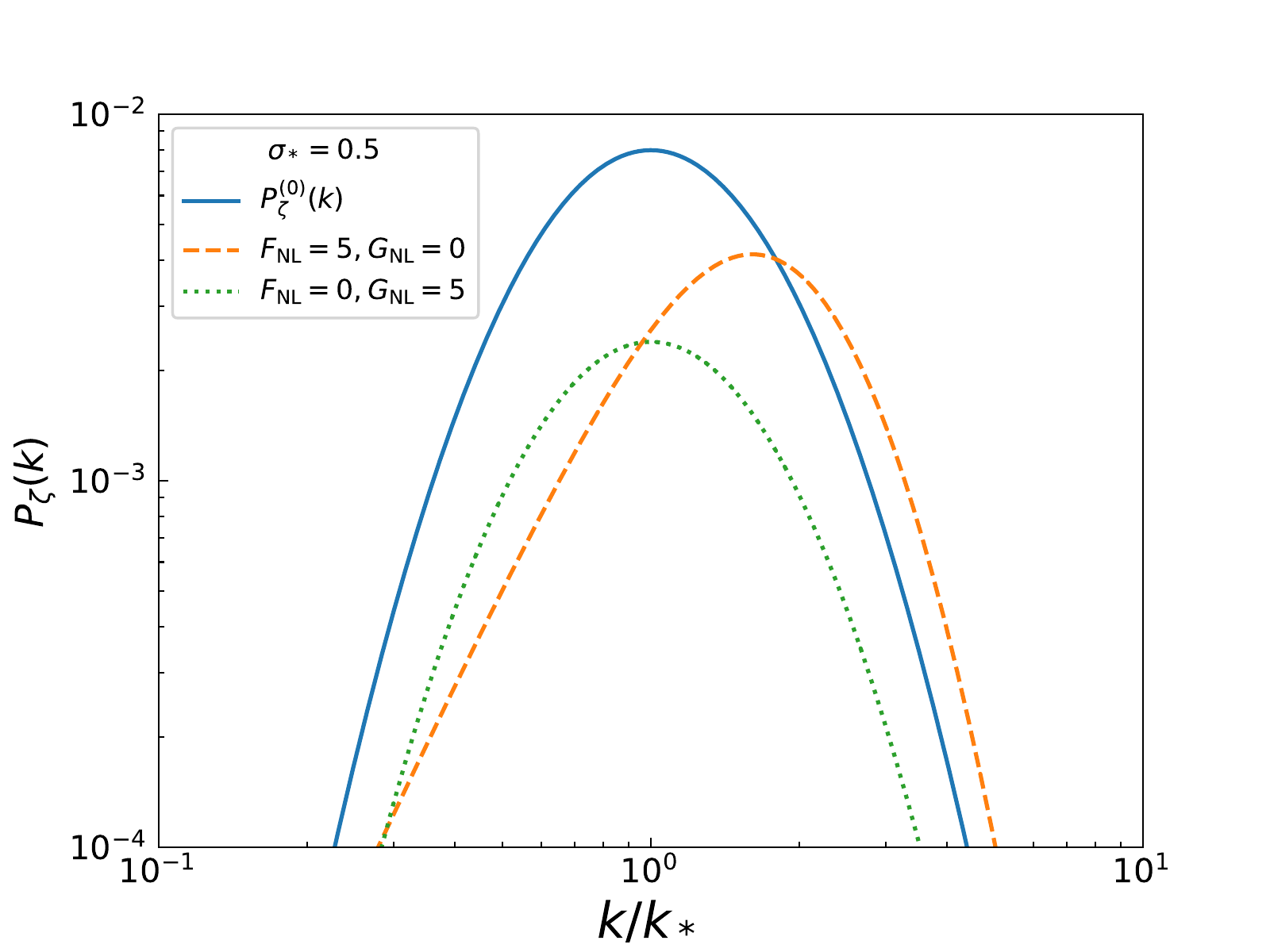}
	\caption{The log-normal power spectrum (the blue solid line) and its one-loop corrections (dashed and dotted lines) of local-type non-Gaussianities with different $\fnl$ and $\gnl$. The width of the Gaussian spectrum and its variance are fixed at $\sigma_* = 0.5$ and $A = 10^{-2}$ respectively.}
	\label{p0p1loc}
\end{figure}

\begin{figure}
    \centering
    \includegraphics[width=0.9\columnwidth]{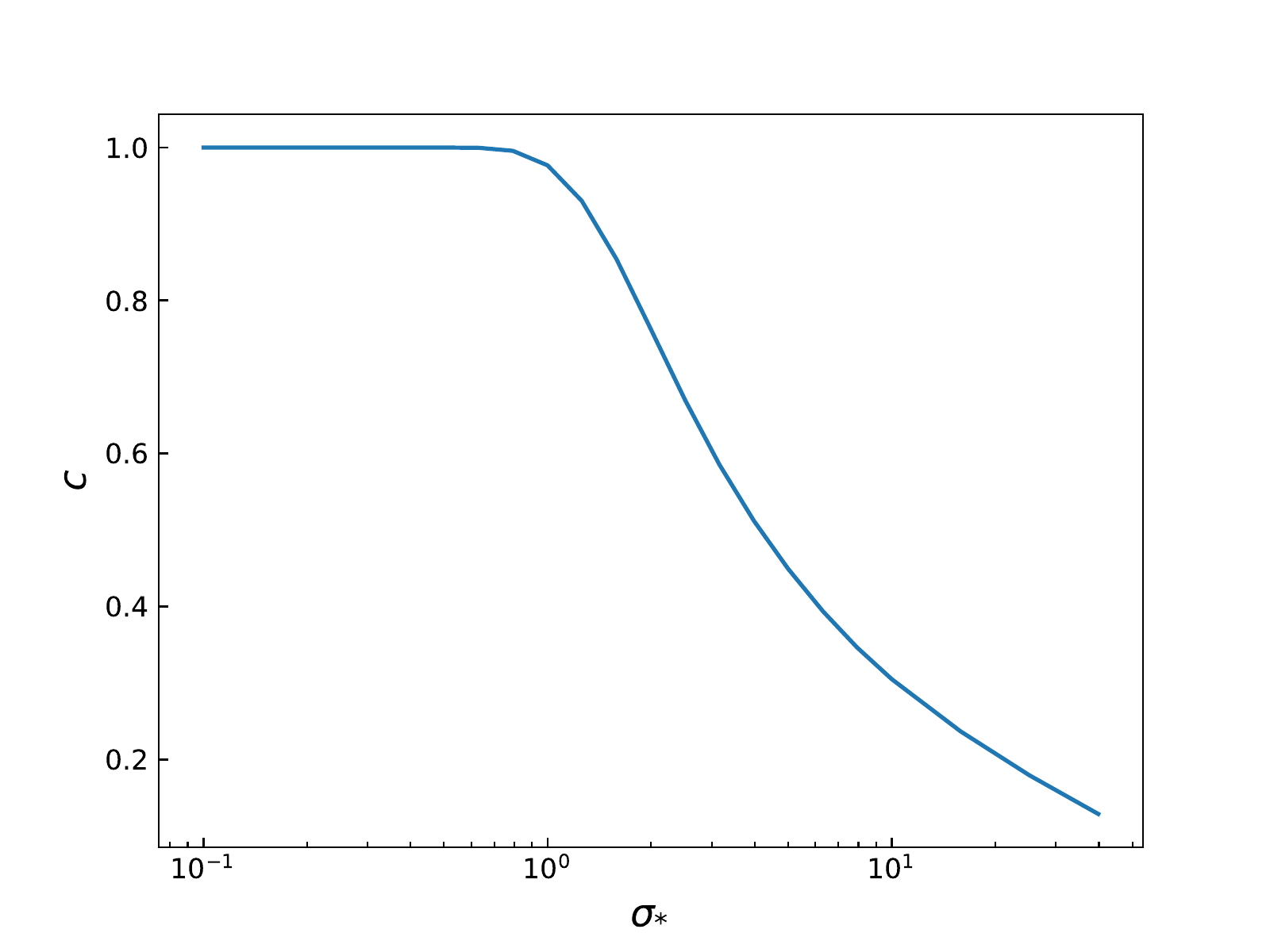}
    \caption{The value of $c$ as a function of the width of the log-normal spectrum $\sigma_*$.}
    \label{c}
\end{figure}

The log-normal shape spectrum is given by
\e
\mathcal{P}_\zeta^{(0)}(k) = \frac{A}{\sqrt{2 \pi \sigma_{*}^{2}}} \exp \(-\frac{\ln^2 \(k / k_{*}\)}{2 \sigma_{*}^{2}}\),
\q
where the dimensionless parameter $\sigma_*$ is related to the the width of the spectrum ($\sim \mathrm{e}^{\sigma_*}$). The total one-loop correction is the sum of the contributions of $\fnl$ and $\gnl$ terms, and depends on the values of $\fnl$ and $\gnl$. The $\gnl$ term in one-loop correction only causes a constant shift $6A\gnl$, while the $\fnl$ term needs to be calculated numerically. The tree level and one-loop power spectrum are showed in \Fig{p0p1loc} for $\fnl=0,\gnl=5$ and $\fnl=5,\gnl=0$, where we set $A=10^{-2}$ and $\sigma_*=0.5$. For the log-normal power spectrum, the coefficient $c$ in Eq.~(\ref{sigma1}) depends on the width of the tree-level spectrum and is shown in Fig.~\ref{c}. We see that the coefficient $c$ is roughly smaller than ${\cal O}(1)$ in spite of the width.

\section{Constraints on the non-Gaussian parameters combining with the abundance of PBHs}
\label{abundance}

In this section, we will briefly review the abundance of PBHs and give the constraints on the non-Gaussian parameters by considering both loop correction and the abundance of PBHs. Throughout this section, we will consider a $\delta$-spectrum described by Eq.~(\ref{delta}).

Let $P(\zeta)$ to be the PDF of $\zeta$, then the initial mass function of PBHs, $\beta$, can be estimated by integrating the PDF over the region $\zeta>\zeta_c$ where $\zeta_c\sim\mathcal{O}(1)$ \cite{Musco:2004ak,Musco:2008hv,Musco:2012au,Harada:2013epa} is the critical value to form a single PBH:
\e\label{beta1}
\beta = \int_{\zeta > \zeta_c} P(\zeta)\mathrm{d} \zeta=\int_{\zeta(\zeta_g) > \zeta_c} \frac{1}{\sqrt{2\pi A}} \exp \( -\frac{\zeta_g^2}{2A} \) \mathrm{d} \zeta_g,
\q
where we have used $\< \zeta_g^2 \>=\int \mathcal{P}_\zeta^{(0)}(k)~ \mathrm{d}\ln k=A$, and $\beta$ is related to the fraction of PBH DM by,  \cite{Nakama:2016gzw}, 
\e\label{beta}
f_{\mathrm{pbh}} \simeq 2.5 \times 10^{8} \beta\(\frac{g_{*}^{\text {form }}}{10.75}\)^{-\frac{1}{4}}\(\frac{m_{\mathrm{pbh}}}{M_{\odot}}\)^{-\frac{1}{2}},
\q
with $g_{*}^{\text{form}}$ and $m_{\mathrm{pbh}}$ the effective degrees of freedom and the mass of PBHs at formation time respectively. A fixed $f_{\mathrm{pbh}}$ would select the value of $A$, $\fnl$ and $\gnl$. In the following part, we consider all DM is in the form of $10^{-12}\Msun$ PBHs, namely $f_{\mathrm{pbh}}=1$, and then $\beta \simeq 7 \times 10^{-15}$.

\begin{figure}
	\centering
	\includegraphics[width=0.9\columnwidth]{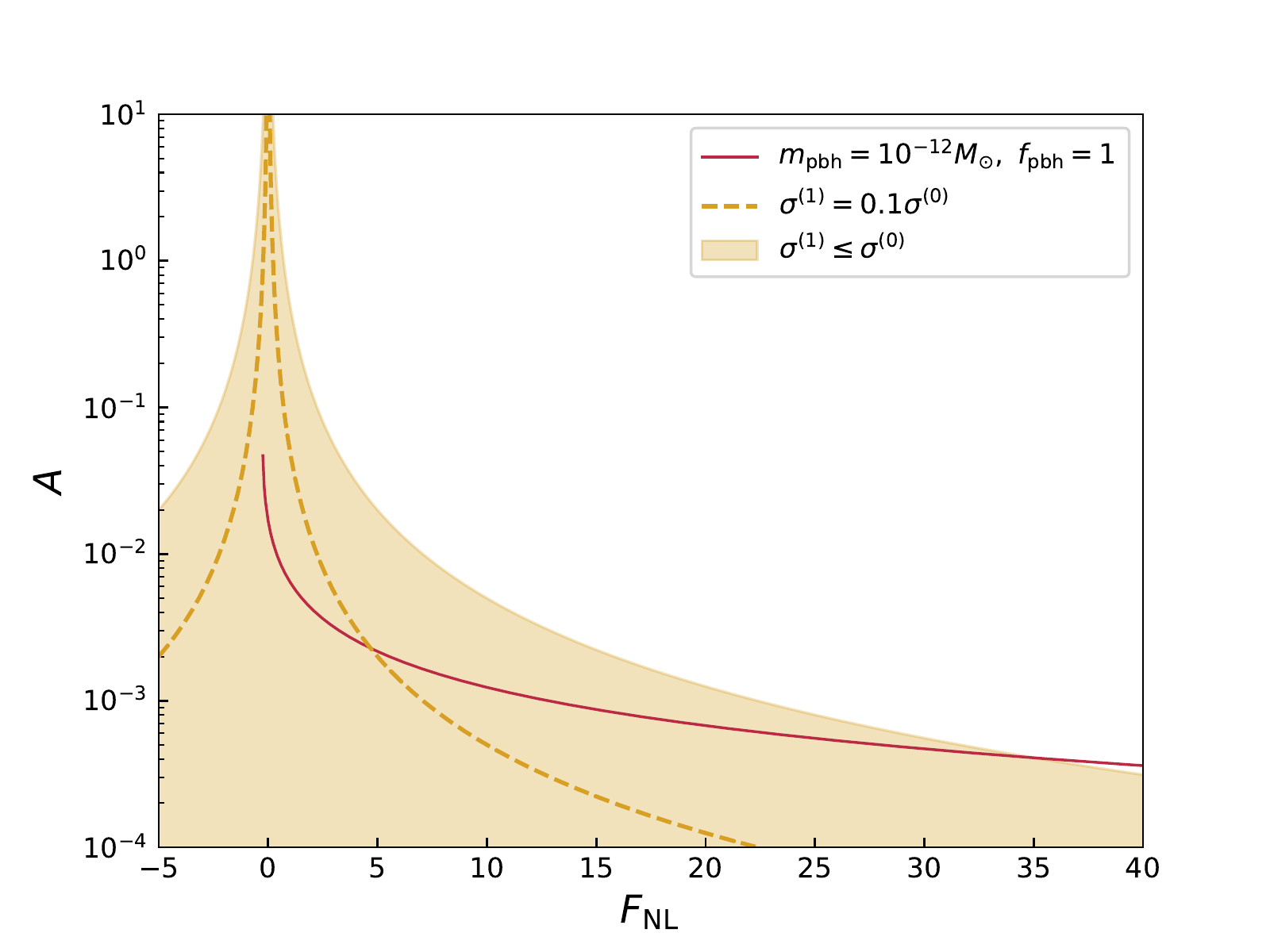}
	\caption{The allowed parameter space for $A$ and $\fnl$ in the case of  $\gnl=0$. The shaded region corresponds to $|\sigma^{(1)}/\sigma^{(0)}|\le1$ and the dashed line corresponds to  $|\sigma^{(1)}/\sigma^{(0)}|= 0.1$. The solid red line stands for $m_{\mathrm{pbh}}=10^{-12} \Msun$ and $f_{\mathrm{pbh}}=1$.}
	\label{af}
\end{figure}

First of all, for a pure $\fnl$ model where $\gnl=0$, equation $\zeta(\zeta_g)=\zeta_c$ is solved as
\e
\zeta_{g\pm} = \frac{-1\pm \sqrt{1+4 \fnl \zeta_c}}{2\fnl}. 
\q
If $\fnl>0$, $\beta$ can be expressed as:
\m
\beta  &&= \int_{-\infty}^{\zeta_{g-}} P(\zeta_g)\mathrm{d} \zeta_g  + \int_{\zeta_{g+}}^{+\infty} P(\zeta_g)\mathrm{d} \zeta_g \no\\
&&= \half\mathrm{erfc} \(\frac{\zeta_{g+}}{\sqrt{2A}}\) + \half \mathrm{erfc} \(-\frac{\zeta_{g-}}{\sqrt{2A}}\),
\n
where $\mathrm{erfc}(x)$ the complementary error function. While for $-\frac{1}{4\zeta_c}<\fnl<0$, $\beta$ becomes
\e
\beta = \int_{\zeta_{g-}}^{\zeta_{g+}} P(\zeta_g)\mathrm{d} \zeta_g = \half\mathrm{erfc} \(\frac{\zeta_{g-}}{\sqrt{2A}}\) - \half \mathrm{erfc} \(\frac{\zeta_{g+}}{\sqrt{2A}}\).
\q
For $\fnl<-\frac{1}{4\zeta_c}$, the curvature perturbation can never exceeds the critical value of forming a PBH.
The parameter space in this case is demonstrated in \Fig{af} (the red solid curve). On the other hand, if we require $|\sigma^{(1)}/\sigma^{(0)}|<1$ to maintain the validity of perturbation theory, this would also place a constraint in the parameter space. The shaded region in \Fig{af} denotes the allowed parameter space which satisfy the perturbativity condition. It can be seen that the two constraints give rise to $-\frac{1}{4}<\fnl\lesssim 35$ for  $|\sigma^{(1)}/\sigma^{(0)}|<1$ and $-\frac{1}{4}<\fnl\lesssim 5$ for $|\sigma^{(1)}/\sigma^{(0)}|<0.1$.

\begin{figure}
	\centering
	\includegraphics[width=0.9\columnwidth]{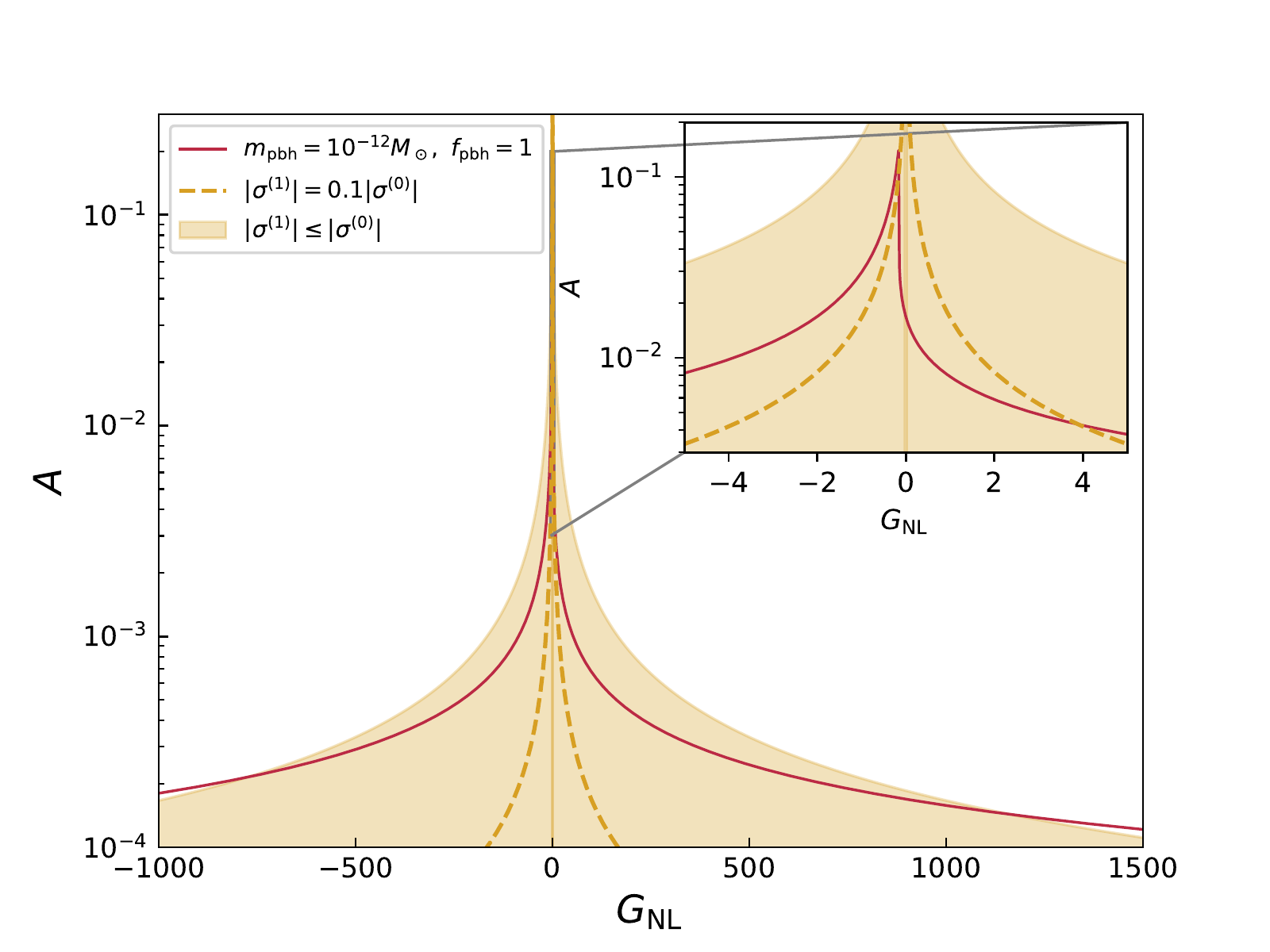}
	\caption{The allowed parameter space for $A$ and $\gnl$ by fixing $\fnl=0$. The shaded region corresponds to $|\sigma^{(1)}/\sigma^{(0)}|\le1$ and the dashed line corresponds to  $|\sigma^{(1)}/\sigma^{(0)}|= 0.1$. If all DM is made up of $10^{-12} \Msun$ PBHs, the choice of $A$ and $\gnl$ falls on the red curve.}
	\label{ag}
\end{figure}

Secondly, we switch to the pure $\gnl$ case where $\fnl=0$. In this case,   $\zeta(\zeta_g)=\zeta_g+\gnl \zeta_g^3=\zeta_c$ has at most three real roots. When $\gnl>0$ or $\gnl<-\frac{4}{27\zeta_c^2}$, there's only one real root, namely
\begin{equation}
\begin{aligned}
\zeta_1 &=-\left(\frac{2^{1 / 3}}{3}\right)\left[\gnl^{2}\left(\zeta_c+\sqrt{\zeta_c^{2}+\frac{4}{27 \gnl}}\right)\right]^{-1 / 3} \\
&+\frac{1}{2^{1 / 3} \gnl}\left[\gnl^{2}\left(\zeta_c+\sqrt{\zeta_c^{2}+\frac{4}{27 \gnl}}\right)\right]^{1 / 3}
\end{aligned}
\end{equation}
and $\beta$ is evaluated as 
\e
\beta = \int_{\zeta_{1}}^{\infty} P(\zeta_g)\mathrm{d} \zeta_g = \half\mathrm{erfc} \(\frac{\zeta_{1}}{\sqrt{2A}}\),
\q
while for positive $\gnl$ it becomes
\e
\beta = \int_{-\infty}^{\zeta_{1}} P(\zeta_g)\mathrm{d} \zeta_g = \half\mathrm{erfc} \(\frac{-\zeta_{1}}{\sqrt{2A}}\),
\q
for $\gnl<-\frac{4}{27\zeta_c^2}$.
When $-\frac{4}{27\zeta_c^2}<\gnl<0$, there are three real roots $\zeta_1<0<\zeta_2<\zeta_3$:
\e
\begin{aligned}
\zeta_1&=-\frac{2}{\sqrt{3}(-\gnl)^{1 / 2}} \cos (\theta / 3) \\
\zeta_2&=\frac{1}{\sqrt{3}(-\gnl)^{1 / 2}}[\cos (\theta / 3)-\sqrt{3} \sin (\theta / 3)] \\
\zeta_3&=\frac{1}{\sqrt{3}(-\gnl)^{1 / 2}}[\cos (\theta / 3)+\sqrt{3} \sin (\theta / 3)],
\end{aligned}
\q
where we used the notations in \cite{Byrnes:2012yx} such that, $\theta=\operatorname{atan}\left[\frac{\left(\zeta_{\mathrm{t}}^{2}-\zeta_{\mathrm{c}}^{2}\right)^{1 / 2}}{\zeta_{\mathrm{c}}}\right]$ and $\zeta_{\mathrm{t}} \equiv \frac{2}{3 \sqrt{3} \sqrt{-\gnl}}$. In this case, $\beta$ takes the form:
\m
\beta  &&= \int_{-\infty}^{\zeta_{1}} P(\zeta_g)\mathrm{d} \zeta_g  + \int_{\zeta_{2}}^{\zeta_3} P(\zeta_g)\mathrm{d} \zeta_g \no\\
&&= \half \mathrm{erfc} \(\frac{-\zeta_{1}}{\sqrt{2A}}\)+\half\mathrm{erfc} \(\frac{\zeta_{3}}{\sqrt{2A}}\) - \half\mathrm{erfc} \(\frac{\zeta_{2}}{\sqrt{2A}}\),\no\\
\n
The parameter space for the pure $\gnl$ case is illustrated in \Fig{ag}. The constraints from PBHs abundance and perturbativity condition lead to $-800<\gnl\lesssim 1200$ for $|\sigma^{(1)}/\sigma^{(0)}|<1$ and $-0.25<\gnl\lesssim 4$ for $|\sigma^{(1)}/\sigma^{(0)}|<0.1$.

\begin{figure}
	\centering
	\includegraphics[width=0.9\columnwidth]{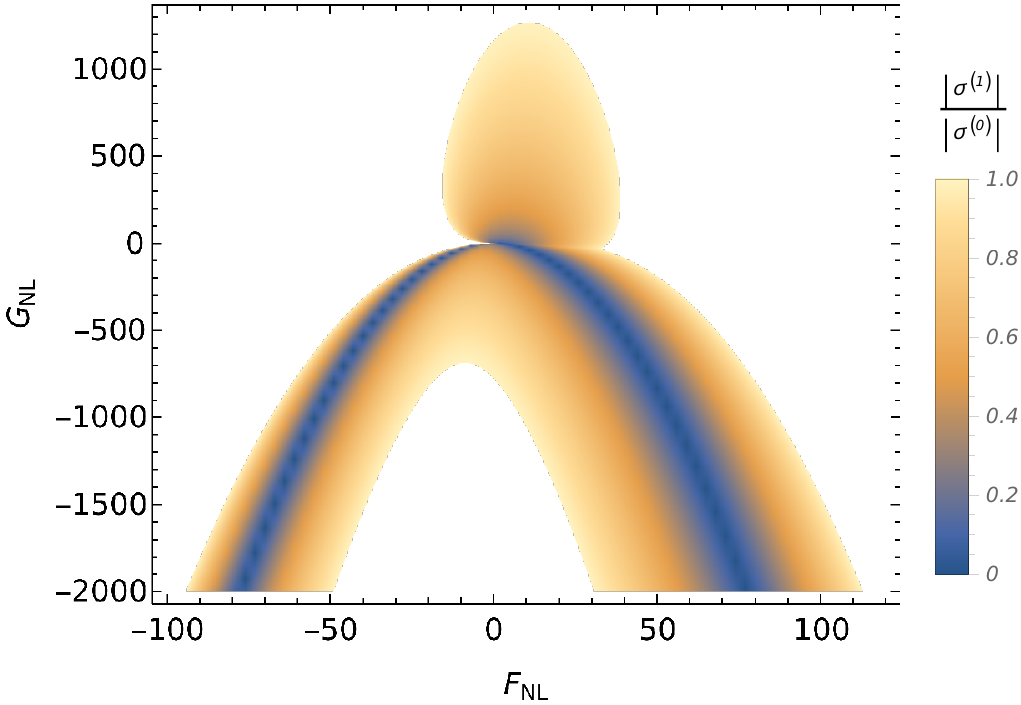}
	\caption{The parameter space for the non-Gaussian parameters $\fnl$ and $\gnl$ if $10^{-12}\Msun$ PBHs make up all of the DM.}
	\label{fnlgnl}
\end{figure}

Finally, for the general case where both $\fnl$ and $\gnl$ are free, the solution to $\zeta(\zeta_g)=\zeta_c$ is lengthy and we calculate $\beta$ numerically. The result is shown in Fig.~\ref{fnlgnl} by fixing $\mpbh=10^{-12}\Msun$ and $\fpbh=1$. 
It can be seen that, in order to maintain the validity of perturbation theory, one can get constraints on both $\fnl$ and $\gnl$ for a fixed $\mpbh$ and $\fpbh$. For $\gnl<0$, the bound of $\fnl$ depends on $\gnl$ and the lower limit of $\gnl$ does not exist and one can only get a constraint on $\fnl$ by considering both perturbativity condition and PBHs abundance.
When $\gnl>0$, one can get constraints such that $-20\lesssim\fnl\lesssim40$ and $\gnl\lesssim 1300$ if $|\sigma^{(1)}/\sigma^{(0)}|<1$, and it turns out that $-1/4\lesssim\fnl\lesssim4$ and $\gnl\lesssim 5$ if $|\sigma^{(1)}/\sigma^{(0)}|<0.1$.

\section{Conclusion and Discussion}
\label{cd}

In this paper, we calculate the one-loop correction to the power spectrum of curvature perturbation with local-type non-Gaussianities. We evaluate one-loop power spectrum to a general form, and take $\delta$ spectrum and log-normal spectrum as two examples. In order to warrant the validity of perturbation theory, we require the variance of the one-loop spectrum is much smaller than that of tree level and we get a perturbativity condition for the  non-Gaussian parameters, namely $|2cA\fnl^2+6A\gnl|\ll1$. Moreover, the non-Gaussian parameters are tightly constrained if a significant amount of DM is in the form of PBHs.

In general, the non-Gaussian parameters of different orders in the local-type non-Gaussian model should be independent of each other, so it is expected to have no accidental cancellation between the $\fnl$ and $\gnl$ terms in the one-loop correction. In this sense, each term in the correction should be respectively smaller than the tree-level order and the relations $A\fnl^2\ll \mathcal{O}(1)$ and $|A\gnl|\ll \mathcal{O}(1)$ should hold. 
On the other hand, the abundance of PBHs would naturally select the non-Gaussian parameters and thus leading to further constraints on the non-Gaussian parameters. For a certain inflation model, the non-Gaussian parameters discussed in this paper could be related to the coefficients in the interaction Hamiltonian above third-order. 
And our work suggests that the consideration of both perturbativity condition and PBHs abundance would place natural constraints on inflation models, which we will leave for future work.

\vspace{1cm}
{\it Acknowledgments. }
We acknowledge the use of HPC Cluster of ITP-CAS. This work is supported by the National Key Research and Development Program of China Grant  No.2020YFC2201502, grants from NSFC (grant No. 11975019, 11991052, 12047503), Key Research Program of Frontier Sciences, CAS, Grant NO. ZDBS-LY-7009, CAS Project for Young Scientists in Basic Research YSBR-006, the Key Research Program of the Chinese Academy of Sciences (Grant NO. XDPB15).

\bibliography{./ref}

\end{document}